\documentclass{vldb}

\usepackage{listings}
\usepackage{framed}
\usepackage{float}
\usepackage{multirow}
\usepackage{url}
\usepackage[T1]{fontenc}
\usepackage{rustic}
\usepackage{times}
\usepackage{courier}
\usepackage{color}

\begin{document}
	
	\newenvironment{packed_item}{
	\begin{list}{$\bullet$}{
	  \setlength{\topsep}{0pt}
	  \setlength{\itemsep}{-1.5pt}
	  \setlength{\parskip}{1pt}
	  \setlength{\labelwidth}{15pt}
	  \setlength{\leftmargin}{10pt}
	  \setlength{\itemindent}{0pt}}
	}{\end{list}}

\title{{\huge MDCC}: Multi-Data Center Consistency}

\numberofauthors{2}
\author{
\begin{tabular}{cccc}\\
Tim Kraska & Gene Pang & Michael J. Franklin & Samuel Madden$^{\spadesuit}$
\end{tabular}
\\
\begin{tabular}{cc}\\
\affaddr{UC Berkeley} & \affaddr{$^{\spadesuit}$MIT} \\
\affaddr{\{kraska, gpang, franklin\}@cs.berkeley.edu} & \affaddr{madden@csail.mit.edu}
\end{tabular}
}

\maketitle
\begin{abstract}
Replicating data across multiple data centers not only allows moving the data closer to the user and, thus, reduces latency for applications, but also increases the availability in the event of a data center failure.
Therefore, it is not surprising that companies like Google, Yahoo, and Netflix already replicate user data across geographically different regions.

However, replication across data centers is expensive.
Inter-data center network delays are in the hundreds of milliseconds and vary significantly.
Synchronous wide-area replication is therefore considered to be unfeasible with strong consistency and current solutions either settle for asynchronous replication which implies the risk of losing data in the event of failures, restrict consistency to small partitions, or give up consistency entirely.
With MDCC (Multi-Data Center Consistency), we describe the first optimistic commit protocol, that does not require a master or partitioning, and is strongly consistent at a cost similar to eventually consistent protocols.
MDCC can commit transactions in a single round-trip across data centers in the normal operational case.
We further propose a new programming model which empowers the application developer to handle longer and unpredictable latencies caused by inter-data center communication.
Our evaluation using the TPC-W benchmark with MDCC deployed across 5 geographically diverse data centers shows that MDCC is able to achieve throughput and latency similar to eventually consistent quorum protocols and that MDCC is able to sustain a data center outage without a significant impact on response times while guaranteeing strong consistency.

\end{abstract}

\section{Introduction}
\label{sec:introduction}

Tolerance to the outage of a single data center is now considered essential for many online services \cite{AppEngine,pnuts,rds_multi_az,AmazonOutage}.
Achieving this in a database system requires replicating data across multiple data centers, and keeping those replicas synchronized and consistent.
For example, Google's e-mail service Gmail synchronously replicates across five data centers to sustain two data center outages: one planned and one unplanned.

\begin{figure}[h*]
 \centering
 \includegraphics[width=\linewidth]{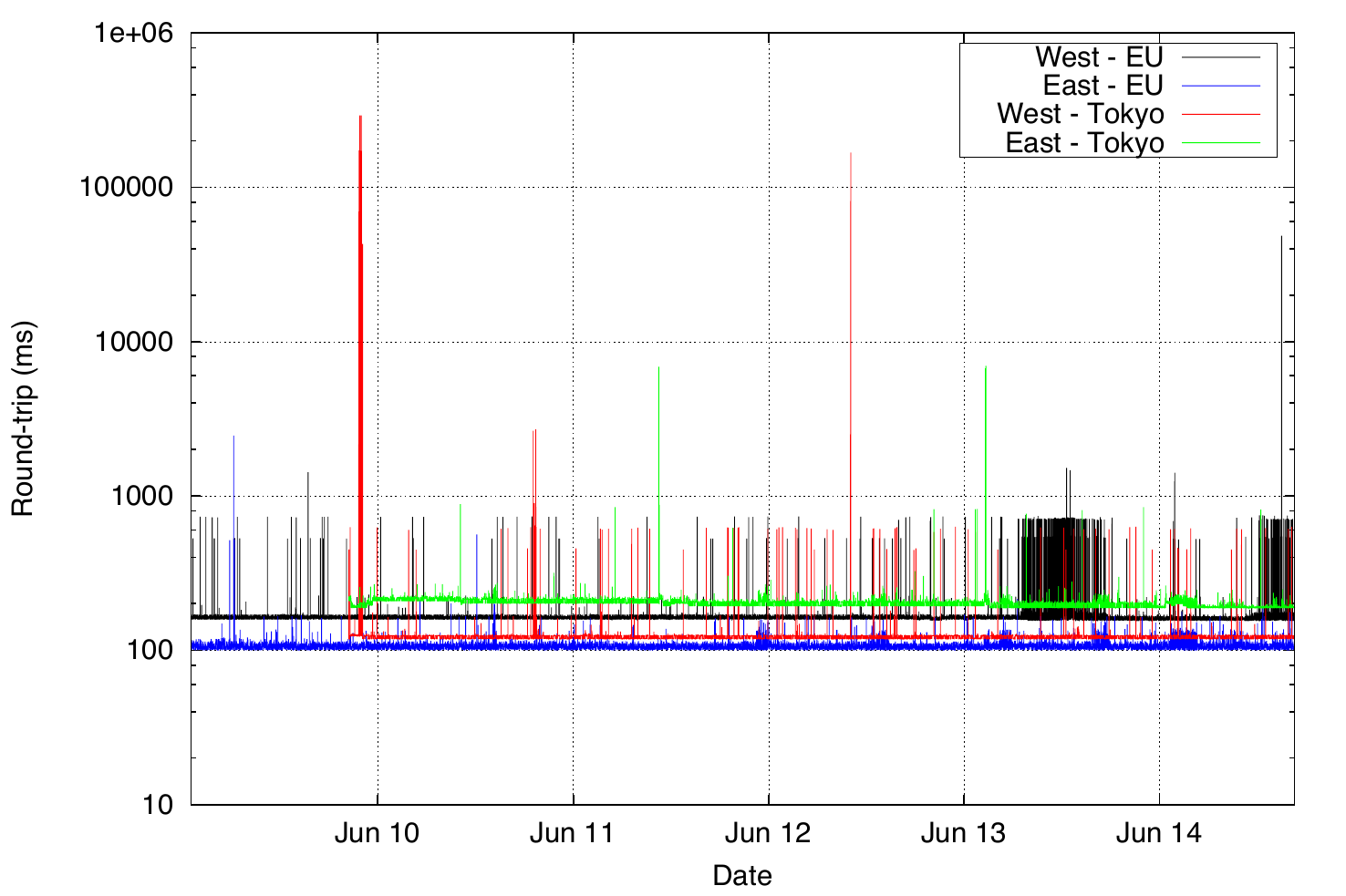}
 \caption{Round trip response times between various regions on Amazon's EC2 cluster.}
 \label{fig:ec2_roundtrips}
\end{figure}

Replication across data centers, however, is expensive.
Inter-data center network delays are in the hundreds of milliseconds and vary significantly as shown in Figure~\ref{fig:ec2_roundtrips} for message delays between different Amazon regions.
Traditional commit protocols for ensuring transactional consistency in distributed databases (e.g., two-phase commit (2PC)) were not designed for these and highly variable latencies that occur in wide area networks.  
First, existing protocols are pessimistic and are required to prepare all resources involved in a transaction by essentially acquiring a lock on the resource, causing an additional message delay and blocking the resource for other concurrent transactions.
These locks are typically held for at least two network round trip times, which, as shown in Figure~\ref{fig:ec2_roundtrips}, can often result in several hundred milliseconds
additional latency.   Such additional latency can significantly impact the usability of websites;  for example~\cite{latency} shows than an additional 200 milliseconds of latency
can result in a significant drop in user satisfaction and ``abandonment'' of websites.

Second, these protocols (2PC in particular) rely on a single coordinator to determine the outcome of a transaction, making them non-resilient to coordinator failure.
In particular, they require locks to be held until the coordinator recovers from a failure. This issue renders
such protocols almost unusable in the wide area, since failures will sometimes occur, and it is not practical to stall transactions until recovery can be completed.
Hence, the traditional ways databases have handled this are either asynchronous replication from a master to one or more replicas (e.g., log shipping) or forfeiting consistency entirely and using an eventually consistent protocol.
In the event of a failure, both approaches may lose committed transactions, become unavailable, or violate consistency.

In this paper, we describe MDCC (short for "Multi-Data Center Consistency"), the first optimistic commit protocol that avoids reliance on a central coordinator and provides strong consistency  at a cost similar to eventually consistent protocols.
Specifically, MDCC requires only a single wide-area message round trip to commit a transaction in the common case, and is "masterless", meaning it can apply reads or updates from any node in any data center.
Similar to 2PC, the MDCC commit protocol can be combined with different read guarantees.
In its default configuration, it guarantees read committed consistency without lost updates by detecting all write-write conflicts.
On the TPC-W benchmark deployed across five Amazon data centers, MDCC reduces per transaction latencies by 50\% (to 234 ms) as compared to 2PC, with transaction throughputs that are twice as high.

In addition to the optimistic commit protocol, MDCC provides a novel programming model that is designed to handle the high variance in round-trip latencies in the wide area, which can result in a small fraction of messages being highly delayed (see Figure~\ref{fig:ec2_roundtrips}).
The programming model is service-level-objective (SLO) aware and allows programmers to give users different responses depending on whether an operation (e.g., a "purchase" in a web store) has committed or is still pending after some time (i.e., SLO).
Specifically, we expose the different stages of a transaction by providing a callback mechanism for the application developer.
This allows writing user-facing applications without sacrificing the user experience in the event of highly delayed messages.

MDCC is not the only system that performs wide-area replication, but it is the only one that provides the combination of low latency (through single message commits) and strong consistency, without requiring a master or a significant database redesign (e.g., the use of static partitions, as in Megastore \cite{Megastore}).
It is the first protocol to use Generalized Paxos \cite{gen_paxos} as a commit protocol, combining it with techniques from the database community (escrow transactions \cite{escrow} and demarcation \cite{demarcation}).
At a high level, the protocol is able to achieve single-message commits by 1) ensuring that every commit has been received by a quorum of replicas, and 2) piggybacking notification of commit state on subsequent transactions.
A number of subtleties had to be addressed to create  a "masterless" approach, including support for commutative updates with value constraints, and to handling conflicts that occur between concurrent transactions.

In summary, the key contributions of MDCC are:

\begin{packed_item}
\item Our new optimistic commit protocol, which achieves wide-area transactional consistency while requiring only 1 network round trip in the common case.
\item  A novel programming model that allows programmers to provide feedback to users about the state of their transactions when there are network or conflict-related commit delays.
\item  Performance results based on the TPC-W benchmark showing that MDCC provides strong consistency with similar costs to eventually consistent protocols. We further show the effects of MDCC's optimizations for the normal operational case, and the performance impact during a simulated data center failure.
\end{packed_item}

The remainder of this paper is organized as follows.
In section \ref{sec:architecture} we show the overall architecture of MDCC.
Section~\ref{sec:programming_model} presents the new programming model that helps developers handle unpredictable and longer network delays between data centers.  
In section \ref{sec:protocol} we describe MDCC's new optimistic commit protocol for the wide area network.
Section \ref{sec:read_consistency} discusses the MDCC's read consistency guarantees.
In section \ref{sec:evaluation} we present our experiments using MDCC across 5 data centers.
Finally, in section \ref{sec:related} we describe related work, and conclude in section \ref{sec:conclusion}.

\vspace{10pt}
\section{Architecture Overview}
\label{sec:architecture}

\begin{figure}[h*]
  \centering
  \includegraphics[trim = 50mm 0mm 50mm 0mm, width=5cm]{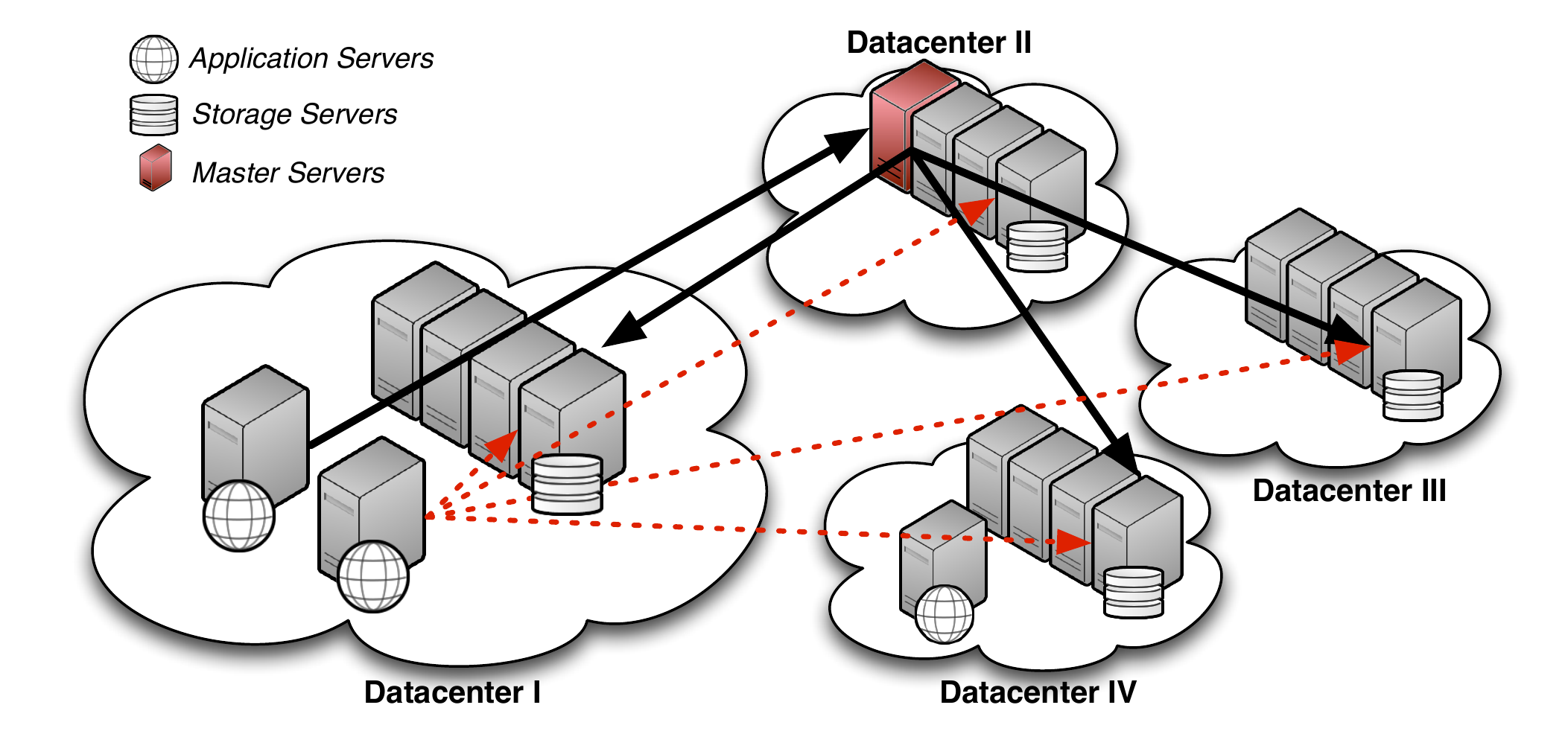}
  \vspace*{-10pt}
  \caption{MDCC architecture}
  \label{fig:arch}
  \vspace*{-5pt}
\end{figure}

MDCC uses a library-centric approach similar to the architectures of DBS3 \cite{db_on_s3} or Megastore \cite{Megastore} (as shown in Figure~\ref{fig:arch}). 
This architecture separates the stateful component of a database system as a distributed record manager. All higher-level functionality (such as query processing and/or transaction management) is provided through a stateless DB library, which can be deployed at the application server. 
As a result, the only stateful component of the architecture, the storage node, is significantly simplified and scalable through standard techniques such as hash partitioning \cite{scads_director}, whereas all higher layers of the database can be replicated freely with the application tier because they are stateless.
Every storage node is responsible for one or more horizontal partitions of the data and partition are completely transparent to the user. 
MDCC places storage nodes in different data centers, which are usually geographically distributed.
Although not required, we assume for the remainder of the paper that every data center contains a full replica of the data, whereas the data itself inside a single data center is partitioned across machines. 

The DB library provides MDCC's programming model for transactions and is mainly responsible for coordinating the replication and consistency of the data throughout the system by implementing MDCC's commit protocol.
In addition, the DB library can either talk directly to the storage servers (red-dotted-arrows in Figure~\ref{fig:arch}) or can choose a storage node to act on its behalf and coordinate the transaction (black-arrows in Figure~\ref{fig:arch}). 
This leads to a very flexible architecture in which storage nodes or application servers can act as the master for a record, depending on the situation (see Section~\ref{sec:protocol}).

In the remainder of this paper, we concentrate on the transaction programming model and the MDCC commit protocol of the architecture.
Other parts of the system, such as load balancing or the storage node design are beyond the scope of this paper and the interested reader is referred to \cite{SCADS, scads_director}.

\section{The MDCC Programming Model}
\label{sec:programming_model}
Because MDCC storage nodes are in different data centers, transactions will need to access data from multiple data centers.
Long and highly variable round-trip latencies between data centers can result in significant transaction execution latencies. 
To help developers cope with these long latencies, MDCC provides a new programming model that allows developers to specify certain callbacks that are executed depending on the different phases of a transaction.

\subsection{State-of-the Art}
Current transaction programming models, such as JDBC or Hibernate, provide little or no support for achieving response-time goals.
Existing programming models implement  a ``fire-and-hope'' paradigm, where once the transaction is executed, the user can only hope that it will finish within the desired time frame.
If the transaction does not return before the application's response-time limit, its outcome is entirely unknown.
In such cases, most applications choose to display a vague message about a server timeout.
 
Listing~\ref{listing:hibernate} shows a ``fire-and-hope'' transaction with Hibernate \cite{Hibernate} the timeout set to 300 milliseconds.
That is, within 300ms the transaction either returns with the outcome stored in the variable {\em success}, or an exception is thrown. 
In the case of an exception, the outcome of the transaction is completely unknown. 


\begin{figure}[t]
\lstset{basicstyle=\small\ttfamily, language=Java, frame=tb, caption=Hibernate Transaction, label=listing:hibernate, captionpos=b}
\begin{lstlisting}
Session sess = factory.openSession();
Transaction tx = null;
try {
  tx = sess.beginTransaction();
  tx.setTimeout(300);
  // The transaction operations
  boolean success = tx.commit();
} catch (RuntimeException e) {
  ...
} finally {
  sess.close();
}
\end{lstlisting}
\vspace*{-20pt}
\end{figure}

Essentially, developers have two options to recover from this unknown state: either they  periodically poll  the database to check if the updates were written, or  ``hack'' the database to get access to the persistent log. 
The first option is almost impossible to implement as it is often unfeasible to distinguish between an application's own changes and changes of other concurrent transactions.
The second option requires detailed knowledge about the internals of the database system and is especially hard in a distributed database system with no centralized log.

\subsection{Language}
MDCC addresses these shortcomings by making service level objectives (SLO) an explicit part of the programming model and by providing a way for applications to register callbacks which are executed depending on the outcome of a transaction.
Furthermore, by making SLOs explicit, the system can take advantage of this information and optimize transaction execution accordingly. 
Code listing~\ref{listing:tx} shows a full example of an MDCC transaction in the Scala programming language.
The details of this listing are explained in the remainder of this section.

\subsubsection{Data Model, Query Language \& Consistency}
\label{sec:programming_model:data_model}
In general, MDCC's programming model can be used with different data models, query languages and consistency guarantees, similar to JDBC being used with SQL or XQuery if the database supports it.
In our current implementation, we provide a simple object-relational data model.
Objects (i.e., tuples) are organized into tables, with the class definition as the schema.
Every class is allowed primary as well as complex types as attributes.
One or more attributes have to be declared as a primary key of the class.
The developer may also annotate attributes with domain constraints, e.g., the {\em stock} attribute must not fall below 0. 
Finally, attributes can be declared to support commutative updates. 
These attributes can then be modified using decrement and increment methods, in addition to the usual getter and setter methods.

MDCC provides a key/value API to retrieve and store objects inside a table, such as \emph{get(key)}, \emph{put(object)}, and \emph{getRange(startKey, endKey)}.
Secondary indexes can be created and probed through a similar interface (i.e., \emph{indexName.get(secondaryKey)}).
MDCC also supports a higher declarative SQL-like query language, called PIQL \cite{PIQL}.
PIQL queries compile down to previously mentioned {\em put, get, getRange} operations and can be statically analyzed for their SLO compliance (see \cite{PIQL}).
For more details on the data and query support, the interested reader is referred to \cite{PIQL}. 

\begin{figure}[t]
\lstset{basicstyle=\small\ttfamily, keywords={onFailure, onAccept, onCommit, finally, finallyRemote}, frame=tb, caption=Full example in Scala, label=listing:tx, captionpos=b}
\begin{lstlisting}
val t = new Tx(300) ({
  // Transaction operations, with get() and put()
  // requests, or PIQL queries
}).onFailure {
  // Error handling code
}.onAccept {
  // Show pending status page
}.onCommit(success => {
  if (success) // Show success page
  else // Show failure page
}).finally( (success, timeout) => {
  // Callback: Update status via AJAX
}).finallyRemote( (success, timeout) => {
  // Callback: Update status via email
})
val status = t.Execute()
\end{lstlisting}
\vspace*{-20pt}
\end{figure}

In our implementation, we allow queries and updates to be interleaved with the application logic. 
However, durability and transactional guarantees are only provided for the table operations (i.e., \emph{put} and \emph{get}). 
That is, as in most database programs, changes to  application variables not stored in the database are immediately visible to concurrent transactions, are not persistent, and are not undone in the event of transaction failure.
As the transaction block executes, it generates a write and a read set of the transaction and all writes are postponed until the end of the transaction. 

MDCC's default consistency guarantee is {\em read committed} without the lost-update problem. 
That is, MDCC guarantees that only committed values are read and prevents all write-write conflicts.
Although we postpone writes until the end of the transaction, we allow during the transaction to read the writes by using a record-cache.
Note, that the default consistency level in most commercial database systems is still the weaker form of read committed \cite{sqlserver_default, oracle_default}, which might lose updates (i.e., overwriting a value from another transaction).
Section~\ref{sec:read_consistency} discusses the other consistency guarantees in more detail and how to achieve higher levels of consistency.

\subsubsection{SLO and Stages}

In the MDCC programming model, developers must explicitly specify a response time SLO for every transaction.
The SLO is a timeout parameter for the transaction; the MDCC system guarantees that execution will return back to the application within the specified SLO time.  
In the example in listing \ref{listing:tx}, the timeout value is 300 milliseconds.  
The programming model forces the developer to explicitly consider the acceptable response times for the end user.

The developer can specify blocks of code for different stages of the transaction. 
These stages of the transaction are \emph{onFailure}, \emph{onAccept}, and \emph{onCommit}. 
The transaction will run and attempt to finish as many stages as possible within the specified timeout.  
At the end of the developer's specified timeout period, the code for latest reached stage is run.  
If nothing is known about the transaction at the timeout, the \emph{onFailure} code is executed.
If the transaction is accepted and is still executing by the database system, but the commit success is unknown when the timeout happens, the \emph{onAccept} code is executed.  
If the transaction finishes before the timeout, the \emph{onCommit} code is executed.
In listing \ref{listing:tx}, the transaction defines the three stages, \emph{onFailure}, \emph{onAccept}, and \emph{onCommit} in order to handle all the possible states of the transaction when the timeout occurs.

Both \emph{onAccept} and \emph{onCommit} are optional.  
There may be situations when the application only needs to know when the transaction was accepted but not yet committed, in which case the developer only needs to supply \emph{onAccept}. 
This means, the transaction will never be lost and will eventual execute, but the application continues without waiting for the commit/abort notification.  
Other applications may only need to know that the transaction completed, in which case the developer will only supply code for \emph{onCommit}.  
The transaction will wait for the commit before returning back to the application.  
If the transaction does not commit before the timeout, then the \emph{onFailure} code will be executed instead.

In addition, there are special callback stages \emph{finally}, and \emph{finallyRemote}.  
The developer provides these callbacks, which  are executed after the transaction has completely executed.  
These callbacks may be executed after the developer-specified timeout, and are used to inform the application of the final outcome of the transaction.  
We describe these clauses next.

\subsubsection{Finally and Finally Remote}
\label{sec:finally}

Both stages, \emph{finally} and \emph{finallyRemote}, are special callback stages used to notify the application of the actual commit decision of the transaction.  
They are different from the other stages because they are not restricted to the specified timeout.  
They are always run after the transaction commits, which may be after the timeout. 
In contrast to the \emph{finally} code which can be anything, the code for \emph{finallyRemote} can only contain web-service invocations (e.g., REST calls), which can be executed anywhere in the system without requiring the outer application context.
For \emph{finally} the system guarantees at-most-once execution. 
For example, a developer might use \emph{finally} to update the web-page dynamically using AJAX about the success of a transaction after the SLO time-out happens. 
However, if the current application server fails, \emph{finally} might be never executed.
In contrast, \emph{finallyRemote} ensures at-least-once execution as the web-service invocation can happen from any service in the system at the cost of reduced expressivity (i.e., only web-service invocations are allowed). 
Listing \ref{listing:tx} shows an example of defining both callbacks, \emph{finally} and \emph{finallyRemote}.

\subsection{Usage Scenarios}
\label{sec:programming_model:uc}

The MDCC programming model is very flexible and can express many kinds of transactions.
A document containing various use cases can be found here \cite{usecases}.
In the following, we describe two use cases in more detail.

\begin{figure}[t]
\lstset{basicstyle=\small\ttfamily, keywords={onFailure, onAccept, onCommit, finally, finallyRemote}, frame=tb, caption=Web shop example, label=listing:tx_amazon, captionpos=b} 
\begin{lstlisting}
val t = new Tx(300) ({
  var order = new Order(cust.key, date)
  orders.put(order)
  var product1 = products.get("Product1")
  var orderline1 = new OrderLine(product1.id, 2)
  orderlines.put(orderline1)
  product1.stock -= 2
  products.put(product1)
}).onFailure {
  // Show error message
}.onAccept {
  // Show page: Thanks for your order!
}.onCommit(success => {
  if (success) // Show success page
  else // Show order not successful page
}).finally( (success, timeout) => {
  if (!timeout) // Update via AJAX
}).finallyRemote( (success, timeout) => {
  // Email user the status
})
\end{lstlisting}
\vspace*{-20pt}
\end{figure}

\vspace{10pt}
\subsubsection{Web Store}

Code listing \ref{listing:tx_amazon} shows an example of a transaction for checking out from a web store such as Amazon.com.
The programming model helps to provide a good user experience independent of the transaction execution time. 
The body of the transaction creates an order, an order line for product1, and reduces the stock of product1.
If the transaction completes in 300ms, the user is immediately informed about the success (or abort) of the order (\emph{onCommit} block).
If the transaction takes longer than 300ms, the user sees a "Thank you for the order" page (\emph{onAccept} block) unless a failure occurred in which case an error message is shown (\emph{onFailure} block). 
In any case, the user will receive an e-mail notification as soon as the transaction completes (\emph{finallyRemote} block).
Furthermore, if the user is still on the "Thank you for the order" page when the transaction finishes, an AJAX call can update the page and show that the order was successful (or failed).

\subsubsection{Twitter}

Code listing \ref{listing:tx_twitter} shows an example of a transaction for sending a tweet.  
In contrast to the previous web shop example, tweets are less critical and are not required to be immediately globally visible. 
Therefore, the transaction only defines the \emph{onFailure} and \emph{onAccept} code blocks, which means the developer is not concerned with the success or failure of the commit. 

\lstset{basicstyle=\small\ttfamily, keywords={onFailure, onAccept, onCommit, finally, finallyRemote}, frame=tb, caption=Twitter example, label=listing:tx_twitter, captionpos=b}
\begin{lstlisting}
val t = new Tx(200) ({
  tweets.put(user.id, tweetText)
}).onFailure {
  // Show error message
}.onAccept {
  // Show tweet accept
}
\end{lstlisting}

\section{The MDCC Protocol}
\label{sec:protocol}
In this section, we describe a new optimistic commit protocol for the wide-area network.
In contrast to pessimistic commit protocols such as two-phase commit, the protocol does not require a {\em prepare phase} and commits updates in a single message round-trip across data centers if no conflicts are detected.
Furthermore, in contrast to recent proposals for low-overhead, single-transaction-at-a-time architectures for single data center deployments (e.g., H-store~\cite{H-Store}), the protocol trades off CPU cycles for fewer message round-trips and increased parallelism, both essential in the wide-area network.

The protocol is based on two key observations of real workloads; either conflicts are rare or updates commute up to a certain limit (e.g.,  a value constraint that the stock should be at least 0).
At its core, the protocol is based on Multi-Paxos, which is able to achieve one-message-round-trip strong consistency for a single key (Section~\ref{sec:protocol:multi-paxos}) using a master per key. 
We extend the protocol to add multi-row transaction support (Section~\ref{sec:protocol:trx}) and show how the protocol works together with the different stages of our programming model (Section~\ref{sec:protocol:stages}).
To this stage the protocol is pessimistic, requires a master per record and thus, also an additional message delay to the master. 
This part of the protocol is the baseline and used whenever the protocol needs to resolve a conflict. 
In Section~\ref{sec:protocol:nomaster} and Section~\ref{sec:protocol:commute}, we extend the protocol and make it optimistic and master free with ideas from Generalized Paxos~\cite{gen_paxos}.
Although Paxos has been used for commit protocols before, we are the first to explore the advanced characteristics of General Paxos for a reliable commit protocol. 

The resulting protocol guarantees read-committed consistency and detects all types of write-write conflicts. 
However, similar to other commit protocols (e.g., 2PC), the protocol can be combined with different read strategies as described in the next Section.

\subsection{Background: Paxos}
\label{sec:protocol:multi-paxos}

In the following we explain the principles of Paxos and how we can use it to update a single record. 

\subsubsection{Classic Paxos}
Paxos is a family of quorum-based protocols for arriving at consensus on a single value among a group of replicas. 
It tolerates a variety of failures including lost, duplicated or reordered messages as well as fail-recovery of nodes. 
Paxos distinguishes between {\em clients}, {\em proposers}, {\em acceptors} and {\em learners}. 
These can be directly mapped to our scenario, where clients are app-servers, proposers are masters, acceptors are storage nodes and all nodes are learners. 
In the remainder of this paper we use the database terminology of clients, masters and storage nodes and assume a master is always one of the storage nodes \footnote{It is also possible to have an app-server as a master to save another  (data center-internal) message.}. 

The basic idea in Classic Paxos is as follows:
App-servers that wish to update a record send their update requests for the record to the master, as shown by the solid lines in Figure~\ref{fig:arch}.
The master informs all storage nodes responsible for the record that he is the master for the next update.
It is possible that more than one master exists, but in order to make progress, eventually only one master per record is allowed. 
The master processes the client request by attempting to convince the storage nodes to {\it agree} on it. 
Storage nodes only agree on an update if it comes from the most recent master they know of and if it has not already agreed to more recent update to the same value.  

The Classic Paxos algorithm \cite{PaxosMadeSimple} operates in two phases.
{\bf Phase~1} tries to establish the mastership for an update for a specific record $r$. 
A master $P$, selects a proposal number $m$, also referred to as ballot number, higher than any proposal number it knows of and sends a \emph{Phase1a} request with the proposal number $m$ to at least a majority of storage nodes responsible for record $r$. 
The proposal numbers are used to determine the latest request and have to be unique.
To ensure uniqueness we concatenate the proposal number with the Server-IP address as the ``real'' proposal number. 
If a storage node receives a \emph{Phase1a} request greater than any other proposal number it has already responded to, it responds with a \emph{Phase1b} message containing the proposal number $m$, the highest-numbered update (if any) including its proposal number $n$, and promises not to accept any \emph{Phase1a} requests less than $m$.
If $P$ receives a response from a majority $Q_C$ of storage nodes containing its \emph{Phase1b} proposal number $m$, it has been chosen as a master. Now, only $P$ will be able to commit a value for proposal number $m$. 

{\bf Phase~2} tries to accept a value. 
$P$ sends an accept request \emph{Phase2a} to each of the storage nodes of Phase 1 with the ballot number $m$ and with a value $v$. 
$v$ is either the update of the highest-numbered proposal among the responses, or it is the requested update from the client if none of the responses contained a value. 
$P$ must re-send the existing update to avoid overwriting it, because a previous Phase~2 may have already learned the existing value with a majority.
If a storage node receives a \emph{Phase2a} request for a proposal numbered $m$, it accepts the proposal, unless it has already responded to a \emph{Phase1a} request having a number greater than $m$, and sends a \emph{Phase2b} message containing the value and ballot number $m$ to the master. 
If the master receives a \emph{Phase2b} message from the majority $Q_C$ of storage nodes for the same ballot number, consensus is reached and the value is learned. 
Afterwards the master informs all other components, app-servers and responsible storage nodes, about the success of the update.\footnote{Actually, it is possible to avoid this delay by sending \emph{Phase2b} messages to all involved nodes. As this increases the number of messages significantly, we do not use this optimization.}

Note, that Classic Paxos is only able to learn a single value per round.
We overcome this limitation by using one separate Paxos round per version of the record with the requirement that the previous version has already been chosen successfully \cite{Megastore, PaxosMadeSimple}.

\subsubsection{Multi Paxos}
The Classic Paxos algorithm requires two message rounds to agree on a value, one in Phase 1 and one in Phase 2. 
If the master is reasonably stable, it is possible to avoid Phase 1 by pre-selecting the master for several rounds \cite{PaxosMadeSimple}.
We explore this by allowing the proposers to suggest the following meta-data in Phase 1:
\begin{verbatim}
[StartRound, EndRound, Ballot, Server-IP]
\end{verbatim}
Thus, the storage nodes can vote on the mastership for several rounds (i.e., version numbers) at once. 
The meta-data also allows for different masters for different versions. 
Storage nodes react to these requests by applying the same semantics for each individual round as defined in \emph{Phase1b}, but answer in a single message. 
The database stores this meta-data including the current version number as part of the record. 
This allows an individual mastership per record. 
To support inserts, MDCC uses a default table-wide meta-data value for round 0 which is stored and modified using the same \emph{Phase1a} messages.
That is, in the default configuration one master per table coordinates all inserts. 
Although a potential bottleneck, the master is normally not in the critical path as it is bypassed as explained at the end of this section. 

\subsection{Transaction Support}
\label{sec:protocol:trx}

The previous subsection showed how we use Multi-Paxos to update a single record. 
We now extend the protocol to support multi-record transactions to achieve  read-committed consistency and avoid lost-updates.
That is, we ensure atomic durability (i.e., either all updates will persist or none) and detect all write-write conflicts. 
We guarantee this consistency level by using a Paxos round per record to accept an {\em option} to execute the update, instead of writing the value directly. 
The transaction is committed and the options are executed only if all the options for every record inside a transactions are successfully learned.

\subsubsection{The Protocol}
For now, we assume that options are physical updates of the form $v_{read} \rightarrow v_{write}$, with $v_{read}$ is the version read by the transaction and $v_{write}$ is the new value. 
This allows us to detect write-write conflicts by testing that the current version of a record is equal to $v_{read}$ of the proposed update.
If they are not equal, the value was modified between the read and write and a write-write conflict was encountered. 
For inserts, we allow an empty $v_{read}$, indicating that a transaction should only succeed if the record doesn't exist. 
Deletes work by marking the item as deleted and are handled as normal updates.
We further allow there to be only one outstanding option per record and that the update is not visible until the option is executed.  

The app-server coordinates the transaction by trying to get the options for all the updates accepted. 
It therefore executes the Paxos protocol as described above for every record in parallel. 
Every storage node responds to the app-server with an accept or reject of the option, depending on if $v_{read}$ differs from the current value or not.
According to the Paxos protocol, the option is learned if and only if a majority agrees on a reject or accept.
That is, we require that even rejected options have to be learned by a majority.
We require further that a new option can only be accepted if the learned option of the previous round was decided. 
This means that all updates to a record will be serialized through the record master. 
Hence, it is impossible for the storage nodes to disagree about a reject or accept for any given option.
We will relax this requirement in Section~\ref{sec:protocol:commute}.

Just as in 2PC, the transaction is successfully committed if every option for every record in the transaction is learned as accepted. 
The transaction is aborted if any of the options is learned as rejected. 
The transaction is undecided as long as there are outstanding options and none of the options learned so far have been rejected. 
It is not possible for a client to manually abort a transaction once it has been proposed. 
This ensures that the status of a transaction depends only on the status of the learned options and hence is always deterministic even in the presence of failures.
If the app-server determines that the transaction is aborted or committed, it informs involved storage nodes through a {\em Learned} message about the decision.  The storage nodes in turn execute the option (commit it) or mark it as rejected. 
Learning an option is always a new Paxos round and thus generates new version-id of the record, regardless of whether the option is accepted or rejected. 

It is important to note that in contrast to 2PC, it is possible to commit the transaction (i.e., commit/abort) in a single round-trip across the data centers if all record masters are local.
This is possible because the commit/abort decision of a transaction depends entirely on the learned values and the application server is not allowed to abort a transaction before at least one option was learned as rejected.
The {\em Learned} message to notify the storage nodes about the abort/commit can be done in the background, does not influence the correctness, and only decreases the abort/wait time between two consecutive updates. 
By adding the transaction support, the protocol is able to achieve 1 round-trip commits if the master is local, but requires two round-trips, one to the master and one from the master to all other storage nodes, if the master is not local.

\subsubsection{Avoiding Deadlocks}
The described protocol is able to atomically commit multi-record transactions. 
It is, however, possible that transactions cause a deadlock by waiting on each others options. 
For example, if two transactions $t_1$ and $t_2$  try to learn an option for the same two records $r_1$ and $r_2$, one transaction might successfully learn the option for $r_1$ and the other transaction for $r_2$.
Recall, that we can not abort the transaction without learning at least one of the options as aborted. 
Hence, both transactions are now deadlocked as each transaction waits for the other option to finish and to learn its own option as aborted or accepted.

We apply a simple pessimistic strategy to avoid deadlocks.
The core idea is to relax the requirement, that we can only learn a new version if the previous round is committed. 
For example, if $t_1$ learns the option $v_0 \rightarrow v_1$ for record $r_1$ in one round as accepted, and $t_2$ tries to acquire an option $v_0 \rightarrow v_2$ for $r_1$, we learn with the next Paxos round both outstanding options $v_0 \rightarrow v_1$ as accepted and $v_0 \rightarrow v_2$ as rejected. 
This simple trick causes transaction $t_1$ to commit and $t_2$ to abort or in the case of the deadlock as described before, both transactions to abort. 

\subsubsection{Failure Scenarios}
Multi-Paxos allows the MDCC commit protocol to recover from various failures. 
For example, a failure of a storage node can be masked by the use of quorums. 
A master failure can be recovered from by selecting a new master (after some timeout) and triggering Phase 1 and 2 as described previously. 
Handling app-server failures is trickier, because an app-server failure can cause a transaction to be forever outstanding as a ``dangling transaction''.
We avoid dangling transactions by including in all of its options a unique transaction-id (e.g., UUIDs) as well as all primary keys of the write-set, and by additionally keeping a log of all learned options at the storage node.
Therefore, every option includes all necessary information to reconstruct the state of the corresponding transactions.
Whenever an app-server failure is detected, e.g., through a timeout on an option by a storage node, the state is reconstructed by reading from a quorum of nodes for every key in the transaction. 
The storage node can then re-execute the transaction as described above. 
Finally, a data center failure is no different than a single node failure.
However, in particular for long outages, it might be useful to develop special bulk-copy techniques to bring the data up-to-date more efficiently without involving the Paxos protocol. 

\subsubsection{Mapping to the Programming Model}
\label{sec:protocol:stages}
The above protocol can be directly mapped to the stages of the programming model in Section~\ref{sec:programming_model}. 
The {\em finally} state is reached when all options of a transaction are learned by the app-server, and the {\em onCommit} state is reached when all options are learned in the SLO timeframe.
The {\em onFailure} state is entered if the app-server does not receive a quorum of \emph{Phase2a} messages within the timeout period, whereas the {\em onAccept} state is entered when at least one \emph{Phase2b} message is received before the SLO timeframe expires. 
This assumes that every storage node will eventually recover and re-execute the transaction. 
Normally, the nearest (local) node will reply the fastest, resulting in the desired short delay to enter the onAccept stage. 
If it is not reasonable to assume that every storage node eventually recovers, the protocol can be configured to wait for more \emph{Phase2b} messages (even from other data centers). 
It is also possible to replicate the transaction id and the write-set to other storage nodes inside the same data center to increase the durability. 

\begin{figure}[h*]
  \centering
  \includegraphics[width=\linewidth]{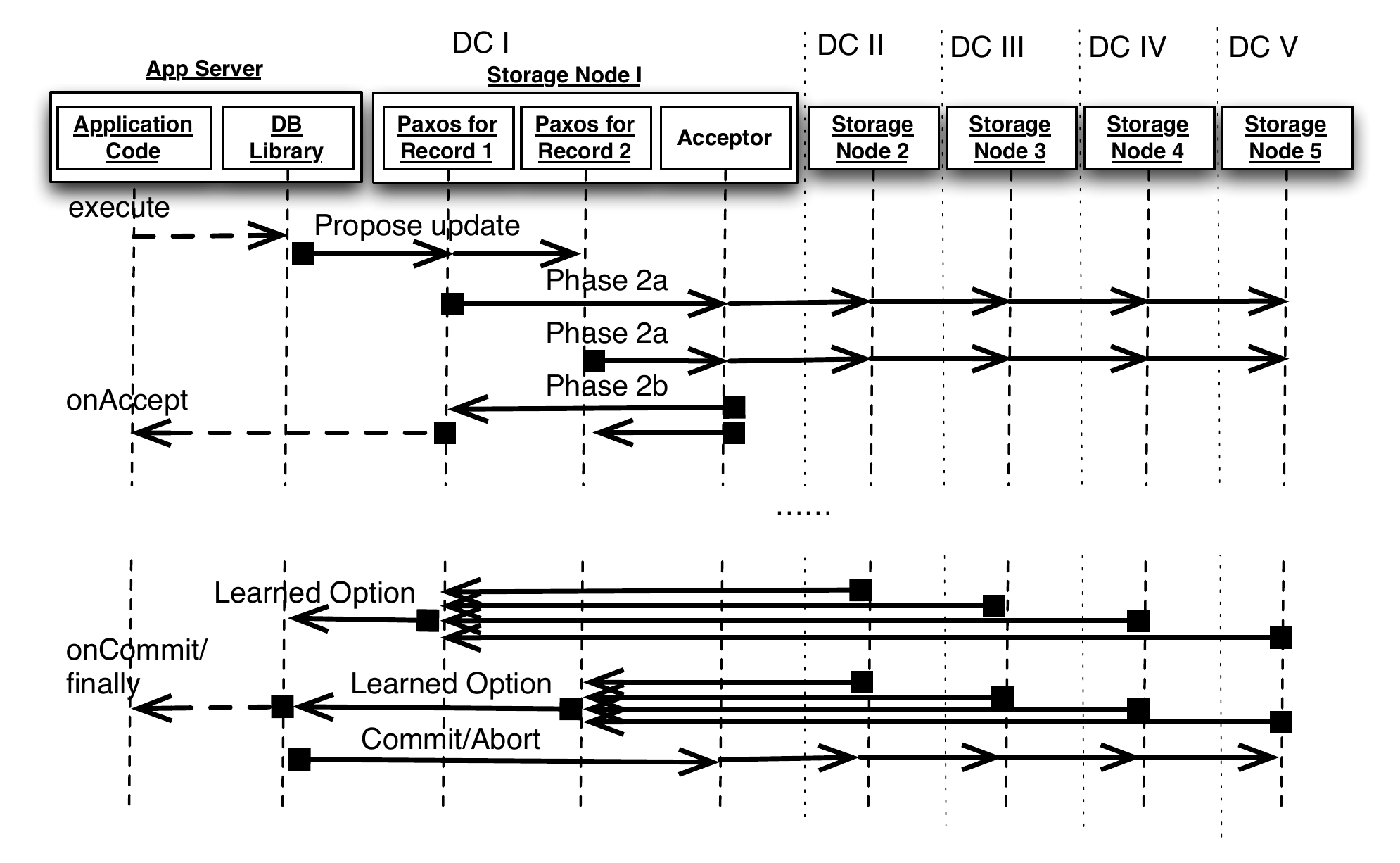}
\vspace*{-10pt}
  \caption{Message Sequence}
\vspace*{-10pt}
  \label{fig:basic_sequence}
\end{figure}

 \subsubsection{Example Execution}
Figure~\ref{fig:basic_sequence} shows the messages involved in a sample execution sequence for a transaction consisting of two records across 5 different data centers.
The state of the overall transaction is managed by the DB library at the application server.
We assume that the local storage node previously established the mastership for the next version for both records.
Therefore, we have two Paxos protocols, one for each record, trying to learn the update as indicated by the two Paxos components inside the storage server. 
The \emph{onAccept} stage is reached after the master has written the \emph{Phase2a} message to stable storage (indicated by the {\em Acceptor} component inside the storage node). 
This assumes that every storage node eventually recovers from a failure as discussed previously.

Although many messages are sent between the data centers, the protocol can determine the result of the transaction in one single-message round trip, as all messages can be sent in parallel. 
After receiving a quorum $Q_C$ of responses per record, the outcome (abort/commit) of the transaction is irrevocably learned, thus allowing the application to move on immediately. 
The {\em Learned} notification to all storage nodes is sent after the application moves on and is only necessary to allow fresh reads without involving a quorum.
We discuss various read strategies in Section~\ref{sec:read_consistency}.

\subsection{Transactions without a Master}
\label{sec:protocol:nomaster}
The previous subsection showed how we achieve transactions with multiple updates in one single round-trip, if the masters for all transaction records are in the same data center as the app-server. 
Still, cases in which the master is in a different data center, or cases in which we need to acquire the mastership, require two round-trips across data centers. 

\subsubsection{Protocol}
Fast Paxos \cite{fast_paxos} avoids the master by distinguishing between {\it classic} and {\it fast} rounds.  
Classic rounds operate like the classic Paxos algorithm described above.
The fast rounds differ from classic rounds in that app-servers send updates directly to storage nodes, bypassing the master.
This saves one message round to the master, which may be in a different data center.
However, since updates are not issued by the master, collisions may occur.
When a collision is detected, the master must step in and resolve it with a classic round.

We adapted the idea of Fast Paxos for MDCC.\footnote{Actually, we did not implement Fast Paxos, but Generalized Paxos which is a super-set of Fast Paxos, as described in the next section.}
In our system all versions start as fast rounds by assuming an implicitly {\em fast} ballot number (e.g., [0,$\infty$,fast=true,ballot=0]), unless the ballot number was explicitly changed by a master through a \emph{Phase1a} message. 
This default ballot number informs the storage nodes to accept all next options from any proposer.

Afterwards, any app-server can propose an option directly to the storage nodes, which in turn promise only to accept the first proposed option.
Simple majority quorums, however, are no longer sufficient to learn a value and ensure safeness of the protocol. 
Instead, learning an option without the master requires a so-called {\em fast} quorum $Q_F$, defined by the following requirements: any two quorums must have a non-empty intersection, and any two fast quorums $Q_F$ and any classic quorum $Q_C$ must have a non-empty intersection.
A typical setting for a replication factor of 5 is a classic quorum size of 3 and a fast quorum size of 4.
If a proposer receives an acknowledgment from a fast quorum, the value is safe and guaranteed to be committed.
However, if a fast quorum cannot be achieved, collision recovery is necessary.

To resolve the collision, a new classic round must be started with \emph{Phase 1}.
After receiving responses from a classic quorum, all potential intersections with a fast quorum must be computed from the responses.
If the intersection consists of all the members having the highest ballot number, and all agree with some option $v$, then $v$ must be proposed next.
Otherwise, no option was previously agreed upon, so any new option can be proposed.
For example, assume the following messages were received as part of a conflict resolution from 4 out of 5 servers with the previously mentioned quorums (notation: (server-id, ballot number, update)) : 
(1,3,$v_0${$\rightarrow$}$v_1$), (2,4,$v_1${$\rightarrow$}$v_2$), (3,4,$v_1${$\rightarrow$}$v_3$ ), (5,4, $v_1${$\rightarrow$}$v_2$)
Here, the intersection size is 2 and the highest ballot number is 4. 
Therefore, the protocol has to compare the following intersections
\begin{eqnarray*} 
&[(2,4, v_1 \rightarrow v_2 ), (3,4, v_1 \rightarrow v_3)]\\
&[(3,4, v_1 \rightarrow v_3 ), (5,4, v_1 \rightarrow v_2)]\\
&[(2,4, v_1 \rightarrow v_2 ), (5,4, v_1 \rightarrow v_2)] 	
\end{eqnarray*}
Only the last intersection has an option in common and all other intersections are empty. 
Hence, the option $v_1${$\rightarrow$}$v_2$ has to be proposed next. 
More details and the correctness proofs of Fast Paxos can be found in \cite{fast_paxos}.

MDCC uses Fast Paxos to bypass the master for accepting an option, which reduces the number of required message rounds.
Per fast round, only one option can be learned. 
However, by combining the idea of Fast Paxos with Multi-Paxos and using the following adjusted ballot-range definitions from Section~\ref{sec:protocol:multi-paxos}, \emph{[StartRound, EndRound, Fast, Ballot, Server-IP]},
it is possible to pre-set several rounds as fast rounds. 
In order to simplify recovery we define further that all classic ballot numbers are always higher ordered than fast ballot numbers.
The rest of the ballot ordering remains as previously defined.
Thus, whenever a conflict is detected, the round is changed to classic, the conflict is resolved and the protocol moves on to the next round, which can either be classic or fast. 
Combined with our earlier definition that a new Paxos round is started only if the previous round is stable and learned, this allows the protocol to execute several consecutive fast rounds without involving a master. 

\subsubsection{Fast-Policy}
\label{sec:prot:fast}
There exists a non-trivial trade-off between fast and classic rounds. 
With fast rounds, two concurrent updates might cause a conflict requiring another two message round-trips for conflict resolution, whereas classic rounds consistently require two round-trips, one to either contact the master or acquire the mastership and one for Phase 2.
Hence, fast rounds should only be used if conflicts are rare. 
MDCC applies a simple policy to automatically adjust the round type at run-time. 

As mentioned before, the default meta-data for all rounds and all records are pre-set to fast with [0,$\infty$,fast=true,ballot=0]. 
As the default meta-data for all records is the same, it does not need to be stored per record. 
A record's meta-data is managed separately, only when conflict resolution is triggered.
The policy is then as follows: 
We first calculate the number of successful fast rounds since the last conflict resolution. 
If at least 4 fast rounds were successful, fast rounds saved at least 1 message since the last conflict resolution.
In this case, we only make the current round classic, to resolve the conflict and continue with fast rounds.
If the last conflict resolution is less than 4 rounds ago, we set the next $\gamma$ rounds to classic (we normally do not change the mastership, just make the rounds classic).
After $\gamma$ transactions, fast rounds are automatically tried again. 

This simple policy yields a self-adjusting execution strategy. 
If conflicts are rare, most transactions will execute in fast mode. 
If conflicts are common, transactions execute in classic mode and after $\gamma$ transactions, a fast round tests the conflict likelihood again.

\subsection{Commutative Updates}
\label{sec:protocol:commute}
Fast Paxos allows many transactions to execute in a single round-trip between data centers.
However, all concurrent updates to a given data item conflict, even though their operations might commute.
MDCC supports commutative updates efficiently by using Generalized Paxos \cite{gen_paxos}, which is a super-set of Fast Paxos.
Furthermore, the MDCC protocol takes advantage of the fact that order is not important with commutative updates, by not enforcing transaction order when possible.

\subsubsection{The Protocol}
\label{sec:protocol:comm_prot}

Generalized Paxos \cite{gen_paxos} uses the same ideas as Fast Paxos but extends it by achieving consensus for a sequence of values, not just a single value.
In Fast Paxos, each round agrees on a single value, but with Generalized Paxos, each round agrees on a sequence of values.
Additionally, Generalized Paxos relaxes the constraint that every acceptor must agree on the same exact sequence.
Since some commands may commute with each other, the acceptors only need to agree on the sequences of values which are compatible with each other.
MDCC utilizes the notion of compatibility to ignore ordering with commutative updates.

Fast commutative rounds are always started by a message from the master.
The master sets the {\em base value}, which is the latest committed value.
Afterwards, any client can propose commutative updates to all storage nodes directly using the same option model as before.
In contrast to the previous section, options now contain commutative updates, which consist of one or more attributes and their respective delta changes (e.g., $stock - 1$).
If a fast quorum $Q_F$ out of $N$ storage nodes accepts the option, the update is committed.
Since Generalized Paxos uses sequences of values and since the updates commute, the acceptors can accept multiple proposals in the same round and the orderings do not have to be equivalent on all storage nodes.
This allows the MDCC protocol to stay in the fast round for longer periods of time, bypassing the master and allowing the commit to happen in one message round.
More details on Generalized Paxos are given in \cite{gen_paxos}.

\subsubsection{Global Constraints}
\label{sec:protocol:constraints}
Generalized Paxos has no notion of integrity constraints, e.g., that the stock of an item must be greater than zero.
We ensure domain integrity constraints by requiring storage nodes to only accept an option if the option would not violate the value constraint under any combination of commit/abort outcomes for previously uncommitted options.
For example, given 5 transactions $t_{1...5}$ (arriving in order), each generating an option [$stock = stock -1$] with the constraint that $stock\ge 0$ and current $stock$ level of $4$, a storage node $s$ will reject $t_5$ even though the first four options may abort.
This definition is analogous to Escrow \cite{escrow} and guarantees correctness even in the presence of aborts and failures.

\begin{figure}[h*]
 \centering
 \includegraphics [width=8.2cm]{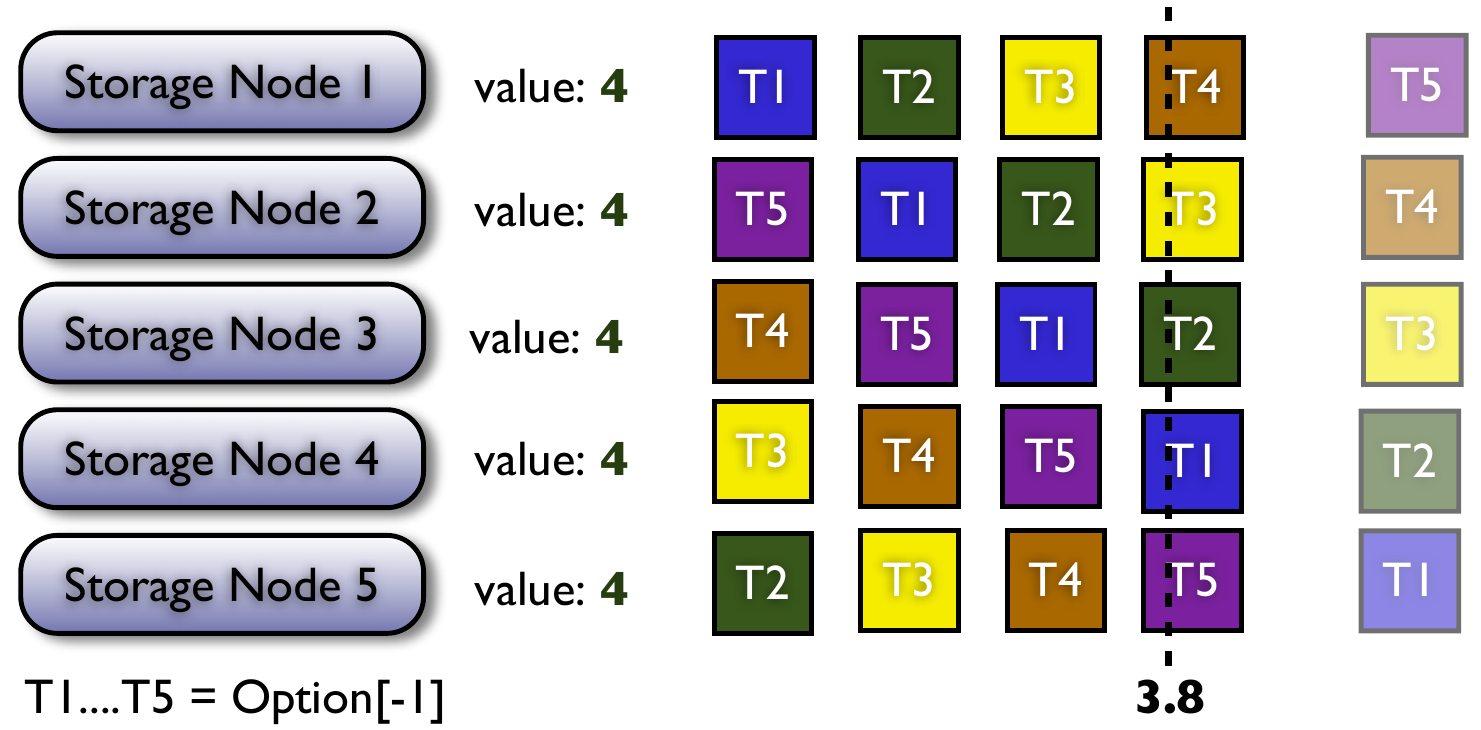}
 \vspace*{-10pt}
 \caption{Message order}
 \label{fig:msg_order}
  \vspace*{-10pt}
\end{figure}

Unfortunately, this still does not guarantee integrity constraints, as storage nodes base decisions on local, not global, knowledge.
Figure~\ref{fig:msg_order} shows a possible message ordering for the above example with 5 storage nodes.
Here, clients wait for 4 (i.e., fast quorum) out 5 responses from nodes, and each storage node makes a decision based on its local state.
Through different message arrival orders it is possible for all 5 transactions to be accepted, even though committing all 5 transactions violates the domain constraint.

We created a new {\it demarcation} strategy \cite{demarcation} for quorum-based systems to avoid this value constraint violation.
Without loss of generality, we now assume a minimum constraint of 0 and that all updates are decrements.
Let $N$ being the replication factor (i.e., the number storage nodes), $X$ be the base value for some attribute and $\delta_i$ be the decrement amount of some transaction $t_i$ on the same attribute. 
If we consider every replicated base value $X$ as a resource, the total number of {\em resources} in the system is $N \cdot X$.
In order to commit an update, $Q_F$ storage nodes must accept the update. 
That is, every successful transaction $t_i$ reduces the resources in the systems by at least $Q_F \cdot \delta_i$.
If we assume $m$ successful transactions with a total decrement amount of $\sum_{i=0}^m{\delta_i} = X$, this means the attribute value reached 0, and the total amount of resources would reduced by at least $Q_F \cdot \sum_{i=0}^m{\delta_i} = Q_F \cdot X$.
Even though the integrity constraint forbids any more decrement transactions on the attribute, it is still possible that the system has $(N - Q_F) \cdot X$ remaining resources due to failures, lost messages or message reordering.

In the worse case, these remaining resources are equally distributed across all the storage nodes. 
Otherwise, at least one of the storage nodes would start to reject options earlier. 
Hence, if we divide the remaining resources $(N - Q_F) \cdot X$ evenly across all the $N$ storage nodes, we can derive a lower limit, which guarantees the value constraint.
Therefore, we require that storage nodes stop accepting options if the option would cause the value to fall below: 
\begin{equation*}
L = \frac{N - Q_F}{N} \cdot X 
\end{equation*}
This limit $L$ is calculated with every new base value and not updated even after receiving a commit for an option. 

At the app-server we detect when options in fast rounds are rejected because of this limit and the protocol handles it as a conflict by switching automatically to classic rounds. 
Again we use the same simple strategy as in the previous sub-section to decide when to switch back to classic rounds with a new limit $L$.

\section{Consistency Guarantees}
\label{sec:read_consistency}
MDCC ensures atomicity (i.e., either all updates in a transaction persist or none) and that two concurrent write-conflicting update transactions do not both commit.
By combining the protocol with different read-strategies it is possible to guarantee various degrees of consistency.

\subsection{Read Committed without Lost Updates}

MDCC's default consistency level is {\em read committed}, but without the lost update problem \cite{sql_isolation_levels}.
The read committed consistency prevents dirty reads, so no transactions will read any other transaction's uncommitted changes.
The lost update problem occurs when transaction $t_1$ first reads a data item $X$, then one or more other transactions write to the same data item $X$, and finally $t_1$ writes to data item $X$.
The updates between the read and write to data item $X$ by $t_1$ are ``lost'' because the final write by $t_1$ overwrites the value and loses the previous updates.
MDCC guarantees read committed consistency, by only reading committed values and not returning the value of uncommitted options.
Lost updates are prevented, as we detect every write-write conflict through the Paxos protocol.

Currently, Microsoft SQL Server, Oracle Database, and PostgreSQL all use read committed consistency as the default level \cite{sqlserver_default, oracle_default, postgres_default}.
We therefore believe that MDCC's default consistency level already suffices for a wide range of applications which might be deployed across data centers.

\subsection{Staleness \& Monotonicity}
Reads can be done from any (local) storage node and are guaranteed to return only committed records.
However, by just reading from a single node, the read might be stale.
For example, if a storage node missed several \emph{Phase2a} messages because of a network problem, reads might return old data.
Reading the latest value requires reading a majority of storage nodes to determine the latest stable version, making reads an expensive operation.

In order to enable guaranteed local up-to-date reads, techniques from Megastore \cite{Megastore} can be applied.
Assuming that the local storage node is the master and that clients are normally co-located in the same data center, a simple strategy is to ensure that the master storage node is always part of the quorum of Phases 1 and 2 (independent of fast or classic rounds), and to remember (locally) whenever this fails.
The same strategy can also be used to guarantee monotonic reads such as {\em repeatable reads} or {\em read your writes}.
That is, whenever we detect that the local data center got outdated, we enforce either quorum reads or change the master storage node to one have the latest updates.
Details on how to support locally consistent reads with one message round-trip time for Paxos are given in \cite{Megastore}.

\subsection{Atomic Visibility}
MDCC provides atomic durability, meaning either all or none of the operations of the transaction are durable, but it does not support atomic visibility.  
That is, some of the updates of a committed transaction might be visible whereas other are not.
One way to achieve atomic visibility is to include full write-set of the transaction with every update.
It is then possible to use standard optimistic concurrency control techniques \cite{JimGrayBook} to detect if a committed transaction was only partially read.
Nevertheless, this strategy for atomic visibility is expensive as it implies storing all write-sets with every update on all records. 

A more lightweight strategy is to partition the data into atomicity blocks which can be stored on a single node.
This enables traditional DBMS \cite{JimGrayBook} locking techniques per storage node to ensure atomic visibility within the partition. 
However, no visibility guarantees are given across partitions. 
The next sub-section describes another strategy using the snapshot isolation protocol.

\subsection{Non-monotonic Snapshot Isolation}
\label{sec:programming_model:nmsi}
MDCC is able to support Non-monotonic Snapshot Isolation (NMSI) {\cite{NMSI}. 
NMSI provides guarantees very similar to SI, while avoiding the overhead of establishing a global transaction order.
In particular with NMSI, transactions are wait-free,  always observe consistent snapshots, and two write-conflicting updates never both commit.
Snapshots taken under NMSI, however, are not totally ordered and strictly consistent. 
For example, NMSI allows to observe the updates of transactions $t_1$ and $t_2$ in one data center, $t_1$ and $t_3$ in another data center and $t_2$ and $t_3$ yet in another data center. 
It was shown in \cite{NMSI}, that no genuine partial replication system can ensure snapshot isolation (SI) as defined in \cite{sql_isolation_levels} or stronger guarantees. 
The MDCC system can be seen as partial replicated system as it enforces no global order, can be used with local reads, and allows commits without involving all replicas.

We are able to achieve NMSI by introducing a local counter per data center. 
This local counter is used to create a per data center consistent snapshot, as proposed by \cite{NMSI} or \cite{walter}.
However, combining MDCC with NMSI is able to improve on the result in \cite{walter} by saving one message for cross data center commits, and improves the protocol of \cite{NMSI} by avoiding the message overhead of tracking complete read and write sets per record.

\section{Evaluation}
\label{sec:evaluation}

We implemented MDCC on top of the key/value store SCADS \cite{SCADS} and evaluated our prototype using the TPC-W and our own micro-benchmark across five different data centers using the Amazon cloud.
To show the benefits of our protocol, we compare the performance and overhead of MDCC to other protocols, namely two-phase commit and eventual consistent quorum protocols.
Furthermore, we show the impact of the different MDCC optimizations on the latency and throughput, and study the latency behavior during a simulated data center failure. 
This section describes the benchmarks, experimental setup, and our findings.

\subsection{Experimental Setup}

 TPC-W is a transactional benchmark that simulates the workload experienced by an e-commerce web server.  
TPC-W defines a total of 14 web interactions (WI), each of which are web page requests that make up of several database queries.  
These WI are requested by emulated browsers (EBs) with a  wait-time between requests, to "realistically" simulate a user interacting with the system. 
TPC-W supports three different user profiles with varying browse-to-buy ratio.
In our experiments, we forego the wait-time between the requests and only use the most write-heavy profile, the ordering-mix, to stress-test our system.
Most of the write transactions consist of either inserts into a table, or commutative updates.
The TPC-W metrics are defined as throughput as Web Interactions Per Second including the page rendering time and total cost of ownership over 3 years.
As our main focus is on the commit protocol, we changed the metrics to response time per transaction and transaction throughput, excluding the page rendering time. 

With our programming model, we implemented the database operations in PIQL \cite{PIQL}, and deployed it across five geographically diverse regions on Amazon EC2: US West (Northern California), US East (Virginia), EU (Ireland), Asia Pacific (Singapore), and Asia Pacific (Tokyo). 
For every region we used two EC2 large instances as storage nodes, which are configured to host a full copy of the data. 
We bulk-loaded 75 emulated browsers' worth of user data for each storage node in the cluster and kept the number of items constant at 10,000.
The comparatively small data size allowed us to better stress-test the different aspects of the MDCC protocol.
If not stated otherwise, we used several EC2 large instances in the US West data center as client machines for the EBs.
Finally, we configured SCADS to use Oracle Berkeley DB (BDB) Java Edition as a persistent storage engine per node instead of its in-memory store.  

\subsection{Comparison with Two-Phase Commit and Quorum Writes}
In the first experiment, we compared the MDCC protocol with two-phase commit (2PC) and quorum writes (QW). 
Two-phase commit is still considered the standard protocol for distributed transactions and operates in two phases: In the first phase, a transaction manager tries to prepare all storage nodes to commit a value.  If all nodes agree to commit the value, 
in the second phase the transaction manager sends a commit to all storage nodes;   otherwise it sends an abort.
Although the protocol looks similar to MDCC, two-phase commit is not fault-tolerant because all storage nodes need to agree on the transaction outcome, and because there is no way to select a new transaction manager.

Quorum writes are the standard for most eventually consistent systems and  are implemented by simply writing the updates to a quorum of storage nodes.
Typically, the majority of the storage nodes is used as a quorum, so that two concurrent writes will have at least one overlapping storage node.
Quorum writes provide no isolation or atomicity guarantees.

\subsubsection{TPC-W Write Response Times}
Figure \ref{fig:tpcw_latencies_cdf} shows the cumulative distribution functions (CDF) of the response times for the different protocols: quorum writes (QW) with a quorum of size 3 and 4 respectively, two-phase commit (2PC), and multi-data center consistency (MDCC).  
We only report the response time for write transaction as read transaction were always local and do not significantly contribute to the latency.
The CDF graph shows that the response times for MDCC are closer to the quorum protocol than to two-phase commit.  
This is not surprising, as we configured MDCC to use fast rounds whenever possible. 
That is, commits are done in one single message across the data centers instead of two with 2PC.
Furthermore, comparing MDCC with quorums writes of sizes 3 and 4, it becomes obvious that MDCC is closer to quorum writes of size 4.
Again, this is no surprise as MDCC's fast quorum require a larger quorum.

It should be noted, however, that a larger quorum of size of 4 as compared to 3 has an influence on the response time. 
This has two reasons: the variance combined with every extra message the protocol has to wait for and the unequal latencies between the data centers. 
Hence, an administrator might configure MDCC to use classic rounds with a local master as the default behavior, if most requests are issued from the same data center.

\begin{figure}[h*]
	\vspace*{-10pt}
  \centering
  \includegraphics[width=\linewidth]{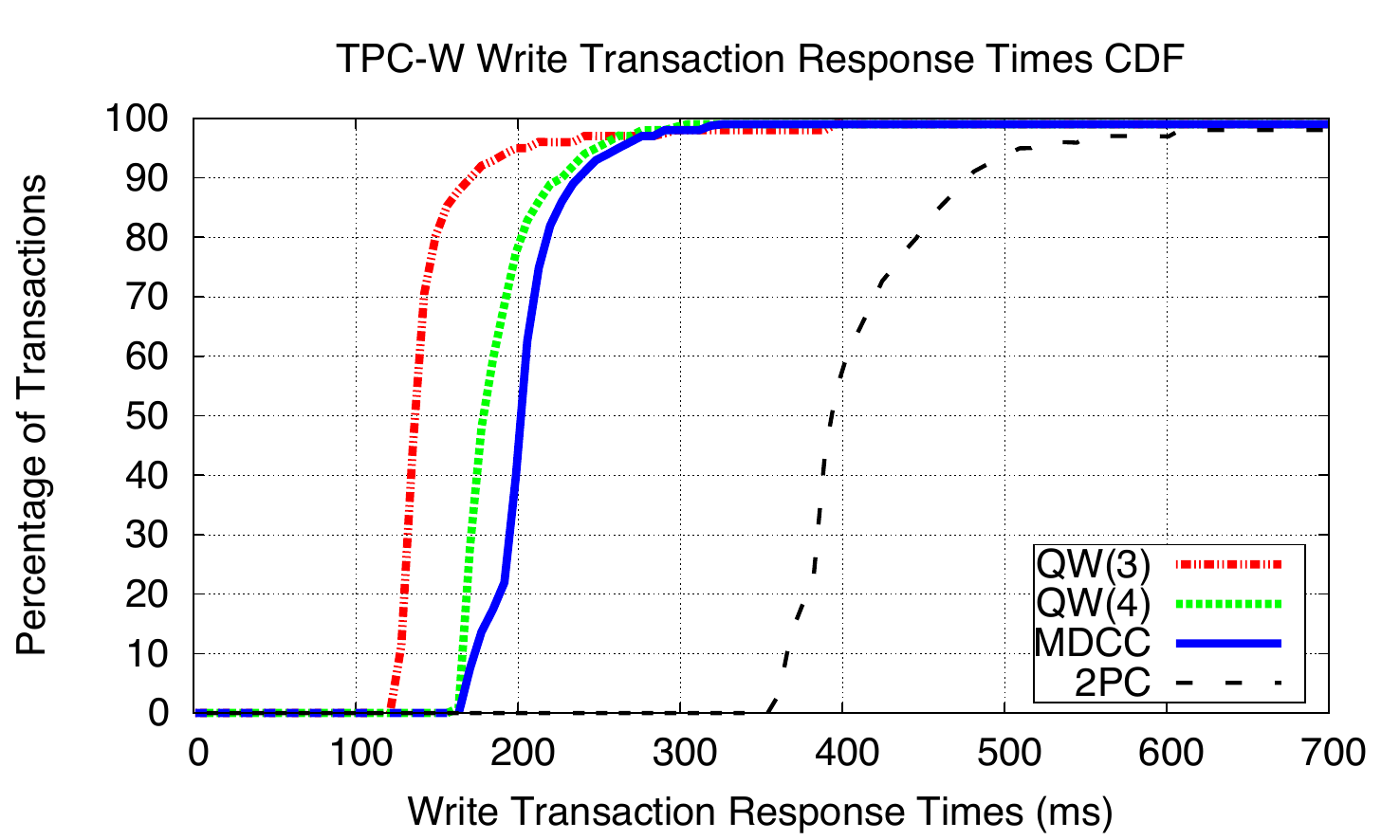}
\vspace*{-20pt}
  \caption{TPC-W write transaction response times for quorum writes (QW), two-phase commit (2PC) and MDCC.}
  \label{fig:tpcw_latencies_cdf}
\vspace*{-5pt}
\end{figure}

\subsubsection{TPC-W Transaction Throughput}
Figure~\ref{fig:tpcw_tps} shows the results of the throughput measurements using the different protocols (excluding quorum 4 for visibility).
In this experiment, we report on read and write transactions for different numbers of concurrent clients (i.e., emulated browsers).
For all protocols, reads were local with a quorum of size 1. 
Figure~\ref{fig:tpcw_tps} shows that the quorum protocol has the lowest overhead and hence the highest throughput, but the MDCC throughput is not far behind.
For 100 concurrent clients, the transaction throughput with the MDCC system was within $10\%$ of the throughput with quorum writes.
The throughput for two-phase commit is significantly lower than the other two protocols, mainly due to the additional wait-time and message overhead for the second round. 

\begin{figure}[h*]
	\vspace*{-5pt}
  \centering
  \includegraphics[width=\linewidth]{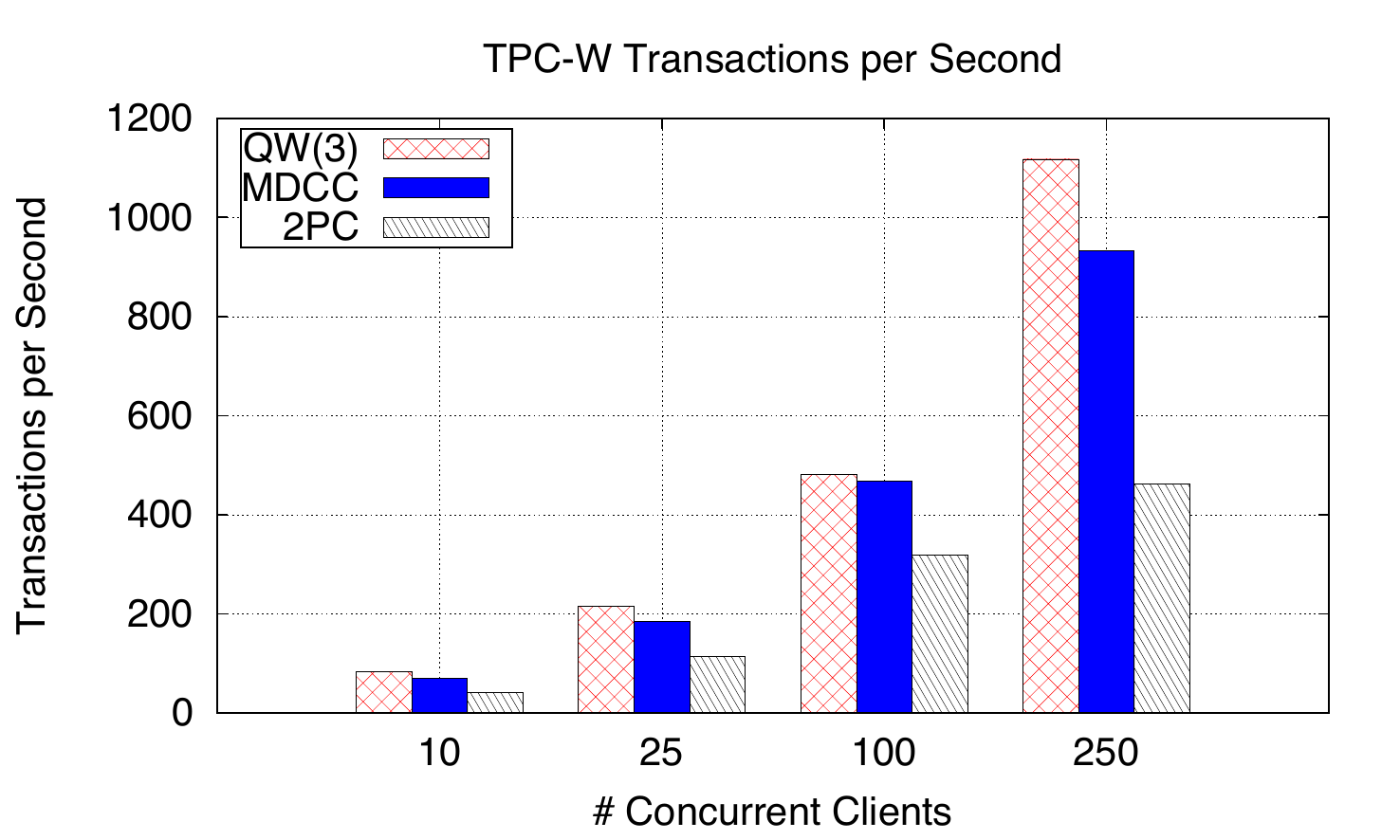}
\vspace*{-20pt}
  \caption{TPC-W transaction throughput for quorum writes (QW), two-phase commit (2PC) and MDCC, and varying number of concurrent clients.}

  \label{fig:tpcw_tps}
\vspace*{-10pt}
\end{figure}

In summary, Figure~\ref{fig:tpcw_latencies_cdf} and ~\ref{fig:tpcw_tps} show that MDCC is able to provide strong consistency for cross data center transactions at a cost similar to eventual consistent protocols. 

\subsection{Micro-Benchmark}
In addition to the TPC-W benchmark, we also used our own micro-benchmark to independently study the different aspects of the MDCC protocol.
We therefore populated a table of 10,000 items, very similar to the TPC-W data. 
We used 100 concurrent clients, each issuing as many write transactions as possible.
All write transactions were of the same type, simulating a purchase of various items.
Each write transaction chose a random number of items to purchase, and then for each item, decremented the stock value by an amount between 1 and 3. 
Each client executed this write transaction on the database as quickly as possible in order to measure the response times and throughput.
For the micro-benchmark, we used the same deployment as for the TPC-W but varied the configurations of the MDCC protocol from only using classic rounds with a master equally distributed across the five data centers (Classic), MDCC using fast rounds but assuming only non-commutative updates (Fast-Non-Comm), and MDCC with fast rounds and commutative updates (Fast-Comm).

\subsubsection{Micro-Benchmark Response Times}
Figure \ref{fig:micro_latency} shows the cumulative distribution functions (CDF) of response times of 2PC and the various MDCC configurations.
We only report on successful transactions. 
The MDCC protocol with classic rounds shows a similar performance as 2PC. 
This can be explained by the fact that we distributed the masters uniformly across the data centers. 
Hence, most of the transactions require an additional message round-trip to contact the master and therefore, also result in two total round-trip times across data centers as 2PC. 

In contrast, both MDCC fast configurations using physical and commutative updates show very similar response times as both require exactly one round-trip time across the data centers for successful transactions. 
However, the response time of physical updates is slightly worse because non-commutative updates cause more message permutations at the different storage nodes to conflict while learning an option and evoke the two extra message delays to resolve the conflict as described in Section~\ref{sec:prot:fast}.
The response time is not significantly worse, because an abort of an transaction does not imply that we also detect a Paxos conflict while learning the option (which requires the two extra delays).
It only means that one of the options is learned as rejected.

\begin{figure}[h*]
	\vspace*{-10pt}
  \centering
  \includegraphics[width=\linewidth]{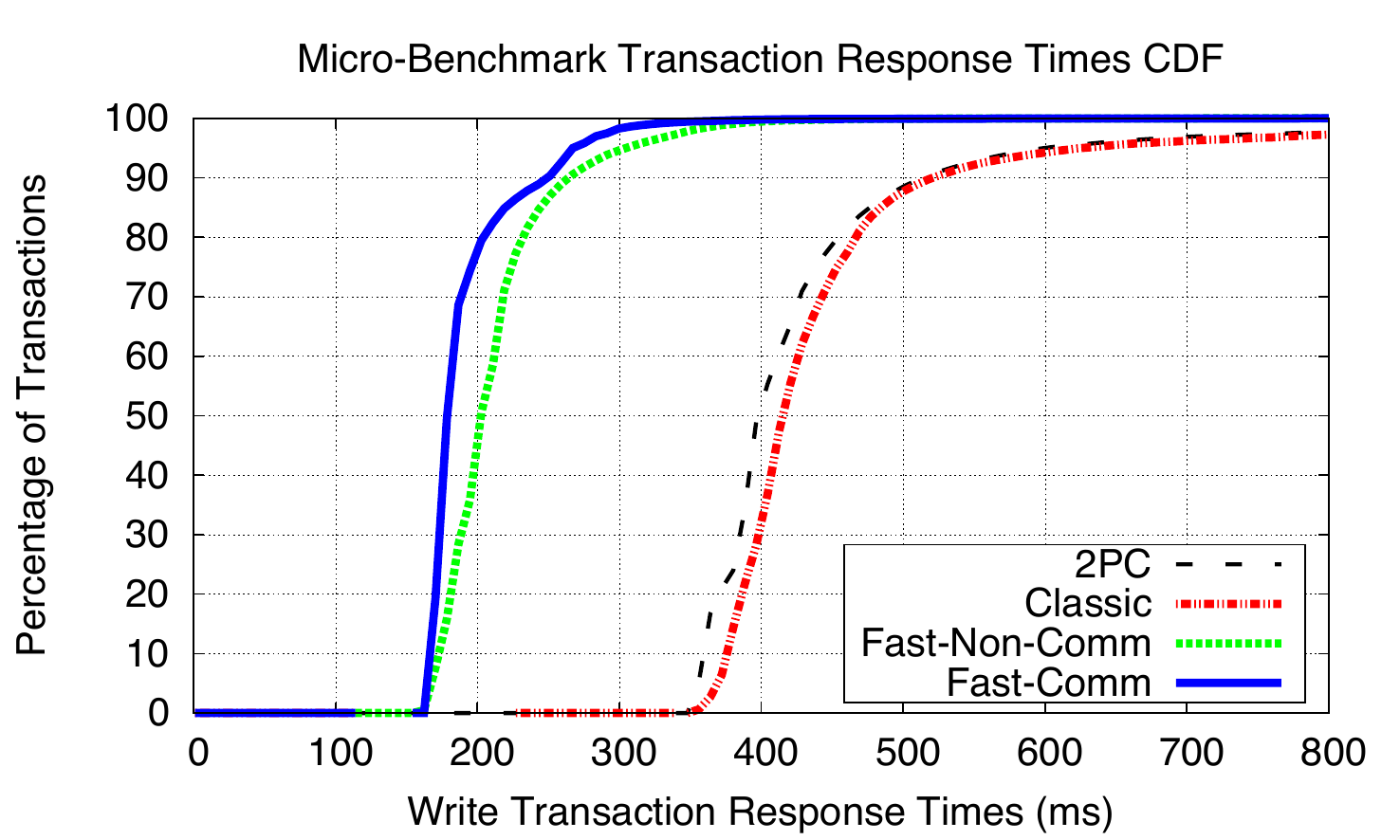}
\vspace*{-20pt}
  \caption{Response times of micro-benchmark transactions for two-phase commit (2PC), and various configurations of MDCC.}
  \label{fig:micro_latency}
\vspace*{-10pt}
\end{figure}

\subsubsection{Micro-Benchmark Throughput}
Figure \ref{fig:micro_tps} shows the transaction throughput for the different protocols and configurations.
Only committed transactions are counted. 
It is clear that the throughput increases with each optimization we enable: from classic to fast rounds, and from non-commutative to commutative updates.
This can be explained by two facts: first, every optimization decreases the conflict likelihood allowing more transactions to be successfully committed, and second, every optimization reduces the number of required messages for conflict resolution, which reduces the overhead. 

\begin{figure}[h*]
\vspace*{-10pt}
  \centering
  \includegraphics[width=\linewidth]{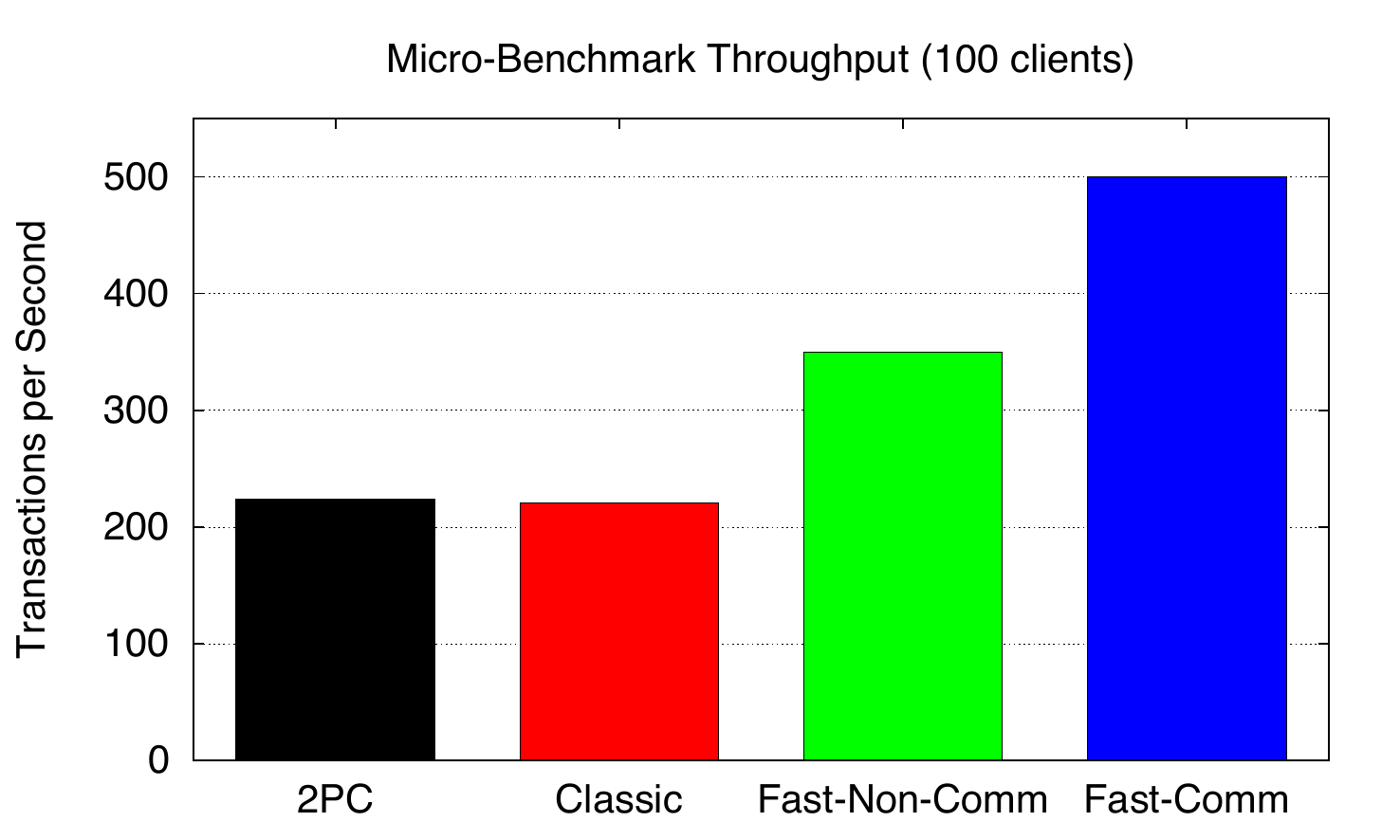}
\vspace*{-20pt}
  \caption{Transaction throughput of the micro-benchmark transactions with 100 concurrent clients.}
  \label{fig:micro_tps}
\vspace*{-10pt}
\end{figure}

\subsubsection{Data Center Fault Tolerance}

We also simulated a full data center outage while running the micro-benchmark.
We started 100 concurrent clients issuing write transactions, from the US-West data center.
About two minutes into the experiment, we simulated a failed US-East data center, which is the data center closest to US-West.
We simulated the failed data center by preventing the data center from receiving any messages.
Since US-East is closest to US-West, it is highly likely that the MDCC commit protocol heavily depends on responses from US-East.
Therefore, ``killing'' US-East forces the protocol to tolerate the failure.
We recorded all the committed transaction response times and plotted the time series graph, in figure \ref{fig:failure_response_times}.

Figure \ref{fig:failure_response_times} shows the transaction response times before and after failing the data center, which occurred at around 125 seconds into the experiment (solid vertical line).
The average response time of transactions before the data center failure was 173.5 ms and the average response time of transactions after the data center failure was 211.7 ms (dotted horizontal lines).
The MDCC system clearly continues to commit transactions seamlessly across the data center failure.
The average transaction latencies increase after the data center failure, but that is expected behavior, since the MDCC commit protocol uses quorums and must wait for responses from another data center farther away.
Waiting for the responses from the next available data center accounts for the increase in average response times after the failure event.
The same argument also explains the increase in variance.
Since fewer servers respond, it is more likely that the protocol is forced to wait for a delayed message. 
These results show MDCC's resilience against data center failures.

\begin{figure}[h*]
  \centering
  \includegraphics[width=\linewidth]{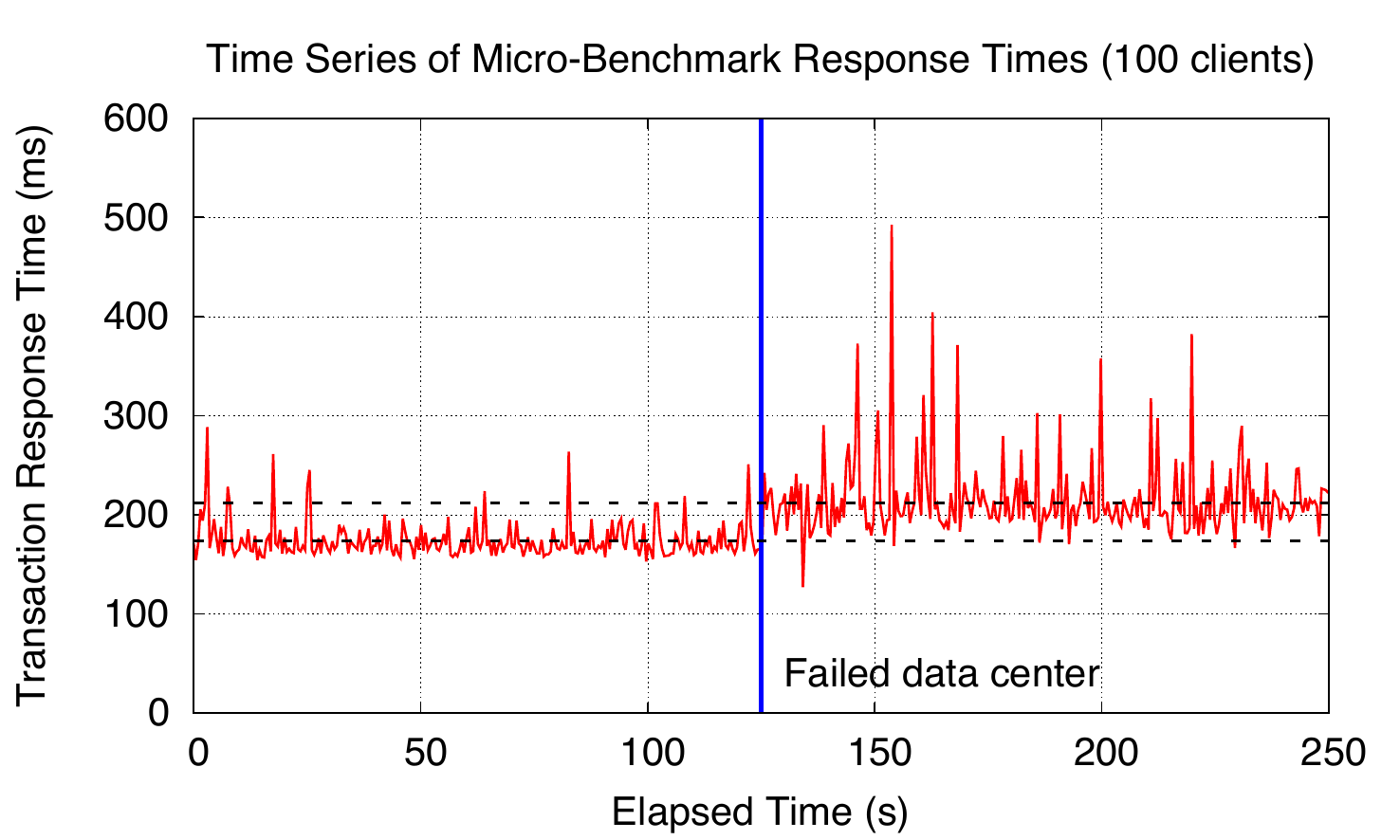}
\vspace*{-20pt}
  \caption{Time series of transaction response times during a data center failure (at 125 seconds).}
  \label{fig:failure_response_times}
\vspace*{-10pt}
\end{figure}

\section{Related Work}
\label{sec:related}

There have been recent interest in scalable datastores replicated across several data centers.  
Megastore \cite{Megastore} is a system which uses Multi-Paxos to synchronously replicate data across multiple data centers (typically five data centers).
Megastore requires to statically partition the data into so called {\em entity groups}. 
Transactional guarantees are only given within a single entity group by using Multi-Paxos to agree on the next change in the entity group.
Hence, all transactions inside a entity group are serialized through a master and may significantly limit the throughput as shown in \cite{cloudbench}.
Across partitions, Megastore uses two-phase commit but it is unclear how failures are handled when transactions span multiple entity groups. 
In contrast, MDCC does not require static partitions and does not involve a master in the normal operational case. 

PNUTS \cite{pnuts} is a geographically replicated datastore which uses asynchronous replication and provides timeline consistency for each record.  
MDCC is also a geographically distributed database, but provides full transactions across records, and uses synchronous replication to prevent data loss from a data center failure.

Amazon's Dynamo \cite{Dynamo} can provide geographically distributed replication of data, but uses a quorum based protocol to achieve eventual consistency.  
MDCC is similar in that, through Paxos, it reads and writes to quorums to tolerate partial failures, but MDCC also provides stronger consistency and multi-key transactions with atomicity guarantees.

Walter \cite{walter} is another geo-replicated system which provides parallel snapshot isolation (PSI) consistency.  
Walter uses asynchronous replication or two-phase commit to propagate changes to remote data centers, so either a data center failure could result in data loss, or two message-rounds are required.
It is actually possible to combine Walter with our commit protocol to achieve PSI with 1-message round commits and without the risk of losing updates in the case of data center failures as described in Section~\ref{sec:programming_model:nmsi}

COPS \cite{cops} provides causal consistency for wide-area storage.
In contrast to MDCC, it does not support transactions and atomicity.
COPS assumes that write-write conflicts can always be resolved (e.g., by using the latest update) and do not cause a set of updates to fail.
MDCC supports transactions and ensures the success of either all or none of the updates in a transaction.
Again, MDCC could be combined to achieve causal consistent reads.

Scalaris \cite{scalaris, atomicCommitmentDHT} is a transactional key-value store built on top of a DHT, and is usually deployed in a geographically diverse environment.  
Scalaris uses Paxos to commit transactions across several keys.  
MDCC similarly uses generalized Paxos to commit transactions, but can take advantage of fast Paxos to bypass the master.

Spinnaker \cite{spinnaker} is a transactional key-value store which uses Paxos for replication.  
Spinnaker focuses on single data center deployments and provides strongly consistent transactions for a single record.  
MDCC focuses on multi-data center replication and supports transactions with multiple records.

The Paxos commit algorithm \cite{consensustrx} shows how to use Multi-Paxos for distributed transactions. 
However, the protocol requires a single master and does not take advantage of rare conflicts or commutative updates.
Furthermore, it also requires to partition the data to resource managers, whereas the MDCC protocol does not require any partitioning scheme.

Finally, many protocols for distributed transactions have been developed \cite{WeikumBook}. 
Our protocol leverages some of the ideas such as the demarcation protocol \cite{demarcation} and to a certain extent, escrow \cite{escrow}. 
However, we present the first optimistic commit protocol which is able to provide read-committed consistency in a single round-trip across all involved nodes.

\section{Conclusion}
\label{sec:conclusion}
The long and fluctuating latencies between data centers make it hard to write highly available applications that can sustain data center failures.
Furthermore, it is still considered unfeasible to achieve synchronized replication across data centers for user-facing applications.
In this paper, we proposed MDCC as a new approach for synchronous replication in the wide-area network.
MDCC consists of two parts: a new programming model and a new commit protocol.
MDCC's transaction programming model provides a clean and simple way for developers to implement user-facing transactions with potentially wildly varying latencies, which occur when replicating across data centers.
MDCC's commit protocol is able to sustain data center failures without compromising availability or consistency at a similar cost as eventually consistent protocols, by requiring only one message round-trip across data centers in the normal operational case.
In contrast to 2PC, MDCC is an optimistic commit protocol and explores the fact that conflicts are rare and/or that updates commute.
It is the first protocol applying the ideas of Generalized Paxos for such optimistic commits.
Furthermore, we described a new policy to optimize the protocol at run-time based on the conflict likelihood and presented the first technique to guarantee value constraints in a quorum-based system.

For the future, we plan to extend our programming model to other languages such as C++, Java, Ruby and .Net to foster adoption in industry.
We also intend to explore more optimizations of the protocol, such as determining the best strategy (fast or classic) based on client locality, or batching strategies to reduce the message overhead.
Finally, combining the commit protocol with other read strategies, in particular PSI, is an interesting future avenue.

MDCC provides the first transaction model and commit protocol for the wide-area network, which is able to achieve strong consistency at a similar cost to eventually consistent protocols in the normal operational case.

\begin{small}
\bibliographystyle{abbrv}
\bibliography{main}
\end{small}
\balancecolumns
\end{document}